\documentclass[aps,pra,reprint,showpacs,superscriptaddress]{revtex4-1}

\usepackage{amsmath}
\usepackage{amssymb}
\usepackage{bm}
\usepackage{setspace}
\usepackage{amsfonts}
\usepackage{rotating}
\usepackage{amsthm}
\usepackage[english]{babel}

\renewcommand{\vec}[1]{\bm{#1}}
\usepackage{graphicx}
\usepackage{epstopdf}

\begin{document}

\title{The quantum mechanical probability density and probability current 
density operators in the Pauli theory}
\author{E.~L.~Rumyantsev}
\author{P.~E.~Kunavin}
\affiliation{Institute of Natural Sciences, Ural Federal University, 620000
Ekaterinburg, Russia}

\begin{abstract}
We present systematic construction of probability and probability current 
densities operators for one-band single particle Pauli equations starting from 
the operators in Dirac electron model within Second Quantized Approach. 
These operators are of importance as in probability interpretation of 
experimental data, so in establishing of boundary conditions. It is shown that 
derived operators differ significally from their convential Schrodinger-type 
counterparts. The generalization of continuity equation for probability density 
under external perturbations and physical meaning of additional source terms is 
discussed. The presented approach can be useful in analysis of carriers dynamics 
described within generic multicomponent  $\vec{k} \cdot \vec{p}$ Hamiltonians in 
Envelope Function Approximation (EFA).
\end{abstract}


\maketitle

\section{Introduction}
The quantum mechanical concept of probability and probability current density is 
of utter importance for interpretation of experimental results connecting the 
wave-like description of the quasiparticle with macroscopic observables 
\cite{PhysRevB.85.235313}. In non-relativistic quantum mechanics the functional 
form of these quantaties, as their physical interpretation, accompanying the 
solution of corresponding Schrodinger equation, are well established. 
Nevertheless, in the general case, the quantum behavior of electron under 
external perturbation must be described by relativistic Dirac equation 
\cite{Rel_Quant_Theory}. The revival of interest in Dirac equation is triggered 
also by recent advances in spintronics and new developments in condense matter 
systems, especially with advent of graphene. It was shown that there exist an 
analogy between the Dirac description of electron kinetics in vacuum and the 
coupled-band formalism for electrons in narrow-gap semiconductors and carbon 
nanotubes \cite{Zawadzki2007}.  At the same time, the relativistic Dirac 
equation poses major open problems which are due to the presence of negative 
energy spectrum interpreted as antimatter (positrons). It is of common consensus 
that the so-called single-particle solutions of Dirac Hamiltonian describe not 
so much particle states as the states with the charge $\pm e$ which are in 
general multiparticle states \cite{Nikishov1986}. Relativistic corrections to 
electron dynamics are taken into account in quantum chemistry calculations, 
especially for systems containing heavy elements \cite{PhysRevA.50.3865}. 
Nevertheless, the relativistic approach based on Dirac-Kohn-Sham equations and 
Dirac-Hartree Fock models \cite{Intro_Rel_Quant_Chem,Rel_Eff_Chem} is too 
complex in general to handle either analytically or numerically 
\cite{PhysRevA.88.032117}. Thus semirelativistic one-band description of 
electron behavior described by so called two-component Pauli equation, which is 
much simpler to solve, is very popular. The problem of approximate reduction of 
four component Dirac problem to two component Pauli problem is solved usially  
by the projection method proposed by Foldy and Wouthuysen (FW) \cite{Foldy1950} 
and its modifications, or by so called elimination method of Pauli 
\cite{Pauli1958}. As a result the electron and positron (hole) degrees of 
freedom are formally disentangled in Dirac Hamiltonian and their dynamics is 
described separatly by two spinor Pauli-like equations. While FW approach is 
without doubt valid for ``free'' Dirac problem, it comes into conflict with 
Quantum Field description in general, when external perturbation is switched on. 
FW approximation ignores the possibility (inevitable in this case) of 
real/virtual pair creation processes. Simultaniously with formulation of 
Pauli-like equations, there arises the question of relativistic corrections to 
Schrodinger probability and current.  Usually, the sought for expressions are 
obtained starting from second-order in $1/c$ Hamiltonian,  obtained through FW 
transformation 
\cite{Hodge2014,PhysRevB.85.235313}.

We intend to show, that straight and physically clear way to derive the form of 
probability and probability current densities operators supplementing 
two-component Pauli-like equations is to view this problem upon within the 
second quantization method (SQM) consided as an alternative to FW-type 
approach. This is not at all new approach to the derivation of Pauli equation 
\cite{Helgaker2006,Helgaker1984,Helgaker1991}. Nevertheless, this approach has 
not been applied for derivation of the form of single particle probability and 
probability current density operators in Pauli theory starting from their 
counterparts in Dirac theory. At the same time, these quantities are crucial 
for formulation of boundary conditions and probability interpretation of 
experimental data. In what follows we adopt the simplest ``no-pair''
approximation, neglecting possible pair production while projecting Dirac 
operators onto single particle channel. 

The paper is organized as follows. In Sec. \ref{sec:2} and Sec. \ref{sec:3} we 
derive the probability density and probability current operators respectively. 
In Sec. \ref{sec:4} we discuss the peculiarities of obtained probability 
continuous equation.

\section{Probability operator}
\label{sec:2}
The four component spinors of plane-wave eigen functions of free problem, 
described by Dirac Hamiltonian \mbox{$H_D = c\vec{\alpha}\vec{k} + \beta mc^2$} 
are chosen for positive energy branch ($\hbar = 1$) in momentum representation 
as
\begin{equation}
 u_{\vec{k},1} = \sqrt{\frac{\varepsilon_k + 1}{2\varepsilon_k}}
  \begin{pmatrix}
   1 \\
   0 \\
   \frac{\lambda_C k_z}{\varepsilon_k + 1} \\
   \frac{\lambda_C k_+}{\varepsilon_k + 1}
  \end{pmatrix},
 u_{\vec{k},2} = \sqrt{\frac{\varepsilon_k + 1}{2\varepsilon_k}}
  \begin{pmatrix}
   0 \\
   1 \\
   \frac{\lambda_C k_-}{\varepsilon_k + 1} \\
   -\frac{\lambda_C k_z}{\varepsilon_k + 1}
  \end{pmatrix},
\end{equation}
and for negative branch the eigen functions are
\begin{equation}
 \tilde{u}_{\vec{k},1} = \sqrt{\frac{\varepsilon_k + 1}{2\varepsilon_k}}
  \begin{pmatrix}
   -\frac{\lambda_C k_z}{\varepsilon_k + 1} \\
   -\frac{\lambda_C k_+}{\varepsilon_k + 1} \\
   1 \\
   0
  \end{pmatrix},
 \tilde{u}_{\vec{k},2} = \sqrt{\frac{\varepsilon_k + 1}{2\varepsilon_k}}
  \begin{pmatrix}
   -\frac{\lambda_C k_-}{\varepsilon_k + 1} \\
   \frac{\lambda_C k_z}{\varepsilon_k + 1} \\
   0 \\
   1
  \end{pmatrix},
\end{equation}
where $\lambda_C = 1/mc$, $\varepsilon_k = \sqrt{1 + \lambda_C^2 k^2}$. We will 
need also ``bare'' plane-wave representation spanned by the following set of 
four-component spinors
\begin{equation}
 u_{\vec{k},1}^0 = 
  \begin{pmatrix}
   1 \\
   0 \\
   0 \\
   0
  \end{pmatrix},
 u_{\vec{k},2}^0 =
  \begin{pmatrix}
   0 \\
   1 \\
   0 \\
   0
  \end{pmatrix},
\end{equation}
\begin{equation}
 \tilde{u}_{\vec{k},1}^0 = 
  \begin{pmatrix}
   0 \\
   0 \\
   1 \\
   0
  \end{pmatrix},
 \tilde{u}_{\vec{k},2}^0 =
  \begin{pmatrix}
   0 \\
   0 \\
   0 \\
   1
  \end{pmatrix},
\end{equation}
Where $u_{\vec{k},1}^0$, $u_{\vec{k},2}^0$ represent ``bare'' electron states 
and $\tilde{u}_{\vec{k},1}^0$, $\tilde{u}_{\vec{k},2}^0$ are ``bare'' hole 
states. This representation is needed as only within this representation in 
position space the momentum operator is obtained by replacing 
$\vec{k}\rightarrow -i\vec{\nabla}$ and position operators  are $c$-numbers. The 
same substitution is used for the solution of multiband $\vec{k} \cdot \vec{p}$ 
Hamiltonians within EFA. Thus the semiconductor problems become analogous to 
the ones in the relativistic theory \cite{PhysRev.148.849}, allowing to extend 
the obtained below results in Dirac theory to the description of carriers in 
semiconductors \cite{2017arXiv170107067Z}.

The field operator can be decomposed into ``dressed'' ($A_{\vec{k},i}$, 
$A_{\vec{k},i}^+$, $B_{\vec{k},i}$, $B_{\vec{k},i}^+$) creation/annihilation 
electron/hole operators, diagonalizing free $H_D$, or into ``bare''  
($a_{\vec{k},i}$, $a_{\vec{k},i}^+$, $b_{\vec{k},i}$, $b_{\vec{k},i}^+$) ones.
\begin{eqnarray}
\nonumber
 \Psi(\vec{r}) &=& 
   \int \left( 
   u_{\vec{k},1}^0 a_{\vec{k},1} + u_{\vec{k},2}^0 a_{\vec{k},2}^+
   + \tilde{u}_{\vec{k},1}^0 a_{\vec{k},1} + \tilde{u}_{\vec{k},2}^0 
a_{\vec{k},2} \right)  \\ \nonumber
&\times& e^{i\vec{k}\vec{r}}d\vec{k} \\
\nonumber
&=&\int \left(u_{\vec{k}, 1}A_{\vec{k},1} + u_{\vec{k}, 2}A_{\vec{k},2}
 + \tilde{u}_{\vec{k},1}B_{\vec{k},1}^+ + 
\tilde{u}_{\vec{k},2}B_{\vec{k},2}^+\right) \\
&\times& e^{i\vec{k}\vec{r}}d\vec{k} \label{eq_field_oper_decomp}
\end{eqnarray}
Using \eqref{eq_field_oper_decomp} it is straightforward to write down 
Bogoluybov transformation from ``bare'' to ``dressed'' operators
\begin{gather}
 \begin{pmatrix}
  a_{\vec{k},1} \\
  a_{\vec{k},2}
 \end{pmatrix}
 = N_k I \begin{pmatrix}
          A_{\vec{k},1} \\
          A_{\vec{k},2}
         \end{pmatrix}
 - M_k \lambda_C \vec{\sigma}\cdot\vec{k} \begin{pmatrix}
                                           B_{\vec{k},1}^+ \\
                                           B_{\vec{k},2}^+
                                          \end{pmatrix}, \\
 \begin{pmatrix}
  b_{\vec{k},1}^+ \\
  b_{\vec{k},2}^+
 \end{pmatrix}
 = N_k I \begin{pmatrix}
          B_{\vec{k},1}^+ \\
          B_{\vec{k},2}^+
         \end{pmatrix}
 + M_k \lambda_C \vec{\sigma}\cdot\vec{k} \begin{pmatrix}
                                           A_{\vec{k},1} \\
                                           A_{\vec{k},2}
                                          \end{pmatrix},
\end{gather}
where
\begin{equation}
 N_k = \sqrt{\frac{\varepsilon_k + 1}{2\varepsilon_k}},\;
 M_k = \frac{1}{\sqrt{2\varepsilon_k(\varepsilon_k + 1)}}.
\end{equation}

The probability density operator $\hat{P}(\vec{R})$ by definition is
\begin{equation}
 \hat{P}(\vec{R}) = \int \Psi(\vec{r})^+ \delta(\vec{R} - \vec{r})
   \Psi(\vec{r}) d\vec{r}.
\end{equation}
Inserting Dirac field operator \eqref{eq_field_oper_decomp} into the latter 
expression we obtain
\begin{eqnarray}
\nonumber
  \hat{P}(\vec{R}) &=& \frac{1}{(2\pi)^3}\sum_{s=1,2}\int\int
   \left( a_{\vec{k}+\vec{q},s}^+a_{\vec{k},s} 
   + b_{\vec{k}+\vec{q},s}b_{\vec{k},s}^+\right) \\
   \nonumber
   &\times& e^{i\vec{q}\vec{R}} d\vec{k} d\vec{q} \\
 &=& \frac{1}{(2\pi)^6} \int \hat{P}(\vec{q}) e^{i\vec{q}\vec{R}} d\vec{q}.
\end{eqnarray}
It must be stressed that position operator is $c$-number only in ``bare''
representation. As it was shown in \cite{2016arXiv161100900R} the position 
operator in dressed representation is non-local operator and its components are 
non-commuting. In what follows we will consider only electron single particle 
channel. Projecting onto positive energy states and neglecting pair 
contribution (stability of vacuum approximation) we obtain in ``dressed'' 
representation
\begin{eqnarray}
 \nonumber
 &\hat{P}_e&(\vec{q}) = \sum_{s=1,2}\int d\vec{k} \Big[
  N_{\vec{k}+\vec{q}}N_{\vec{k}}A_{\vec{k}+\vec{q},s}^+A_{\vec{k},s} \\
  \nonumber &+& M_{\vec{k}+\vec{q}}M_{\vec{k}} \lambda_C^2
   \begin{pmatrix}
     A_{\vec{k}+\vec{q},1}^+ \\
     A_{\vec{k}+\vec{q},2}^+
   \end{pmatrix}^T
   \left( (\vec{k}+\vec{q}) \cdot \vec{\sigma} \right)(\vec{k}\cdot\vec{\sigma})
   \begin{pmatrix}
     A_{\vec{k},1}\\
     A_{\vec{k},2}
   \end{pmatrix} \Big]\\
\end{eqnarray}
Using identity for Pauli matrices $\sigma_i\sigma_j = I\delta_{ij} + 
i\varepsilon_{ijk}\sigma_k$, we can separate the spin dependent terms in this 
expression
\begin{eqnarray}
 \nonumber
 \hat{P}_e(\vec{q}) &=& \sum_{s=1,2}\int d\vec{k} \Big[
  N_{\vec{k}+\vec{q}}N_{\vec{k}} \\ \nonumber
  &+& M_{\vec{k}+\vec{q}}M_{\vec{k}} \lambda_C^2
    \left( (\vec{k}+\vec{q}) \cdot \vec{k} \right) \Big] 
 A_{\vec{k}+\vec{q},s}^+A_{\vec{k},s} \\
 &+& i \lambda_C^2 \int d\vec{k} M_{\vec{k}+\vec{q}}M_{\vec{k}}
    \begin{pmatrix}
     A_{\vec{k}+\vec{q},1}^+ \\ \nonumber
     A_{\vec{k}+\vec{q},2}^+
   \end{pmatrix}^T
   \vec{\sigma}\cdot[\vec{q}\times\vec{k}]
      \begin{pmatrix}
     A_{\vec{k},1}\\
     A_{\vec{k},2}
   \end{pmatrix}. \\ \label{eq_Pe}
\end{eqnarray}
Such decomposition of expression for $\hat{P}_e(\vec{q})$ is a 
semi-relativistic analog of the famous Gordon decomposition of Dirac current 
\cite{Gordon1928}.

In gapless case ($\lambda_C \rightarrow \infty$ or $m\rightarrow 0$), which can 
be of interest while analyzing the behavior of carriers in quazi-relativistic 
``graphene'' case, this expression simplifies to
\begin{eqnarray}
 \nonumber
\hat{P}_e(\vec{q}) = \frac{1}{2}\int &d\vec{k}&
    \begin{pmatrix}
     A_{\vec{k}+\vec{q},1}^+ \\ \label{eq_Prop_oper_graphene}
     A_{\vec{k}+\vec{q},2}^+
   \end{pmatrix}^T 
   \Big[ 1 + (\vec{n}_{\vec{k}} \cdot \vec{n}_{\vec{k}+\vec{q}}) \\
   &+& i\vec{\sigma}\cdot[\vec{n}_{\vec{k}} \times \vec{n}_{\vec{k}+\vec{q}}]
   \Big]
   \begin{pmatrix}
     A_{\vec{k},1}\\
     A_{\vec{k},2}
   \end{pmatrix},
\end{eqnarray}
where $\vec{n}_{\vec{k}} = \vec{k}/|\vec{k}|$. It is seen from 
\eqref{eq_Prop_oper_graphene} that probability of electron scattering from 
the state $|\vec{k}\rangle$ to the state $|-\vec{k}\rangle$ 
($\vec{q}=-2\vec{k}$) is zero. This is manifestation of the 
well-established fact that backscattering is absent in 3D Weyl or 2D graphene 
\cite{Ando1998} case. It must be stressed that this result does not depend on 
the form of external perturbation. This is an inherent property of the defined 
probability operator and is a consequence of introducing vacuum (filled valence 
band in $\vec{k} \cdot \vec{p}$ theory) into the Dirac theory. Note, that this 
result is valid in our approach only if we impose the condition of vacuum 
stability. It must be stressed that in the gapless case the separation of 
modified potential in Pauli equation into ``bare'' and SO parts is 
meaningless. As it is seen from \eqref{eq_Prop_oper_graphene}, all three terms,
including ``bare'' potential term, are of the same order and do not depend on 
the parameters of the problem, but only on the strength of external potential.

As we intend to apply proposed approach elsewhere for constructing density and 
current operators in semiconductors described within multi-component  $\vec{k} 
\cdot \vec{p}$ Hamiltonians, it is of interest to consider low energy limit 
$\lambda_C k \ll 1$. In $\vec{k} \cdot \vec{p}$ theory such approximation is 
known as EFA, while in Dirac theory it is called $1/c$ approximation 
\cite{Rel_Quant_Theory} or $v/c$ approximation \cite{2017arXiv170107067Z}.
As it will be seen this approximation not only simplifies the derived 
expressions but allows the elucidation of the physical meaning of various terms 
entering obtained expressions. Using the approximate expressions for 
$N_{\vec{k}}$ and  $M_{\vec{k}}$ up to the second order in $\lambda_C k$
\begin{gather}
 N_{\vec{k}} \approx 1 - \frac{1}{8}\lambda_C^2 k^2, \\
M_{\vec{k}} \approx \frac{1}{2}\left( 1 - \frac{3}{8}\lambda_C^2 k^2 \right),\\
 \varepsilon_k \approx 1 + \frac{1}{2}\lambda_C^2 k^2
  - \frac{1}{8}(\lambda_C^2 k^2)^2,
\end{gather}
the expression for $\hat{P}_e(\vec{q})$ simplifies to
\begin{eqnarray}
 \nonumber
 \hat{P}_e(\vec{q}) &=& \sum_{s=1,2}\int d\vec{k}
  \Big[ 1 - \frac{1}{8}\lambda_C^2 q^2 \Big]A_{\vec{k}+\vec{q},s}^+A_{\vec{k},s}
  \\ \nonumber
 &+& i\frac{\lambda_C^2}{4}\int d\vec{k}
   \begin{pmatrix}
     A_{\vec{k}+\vec{q},1}^+ \\ \label{eq_Pe_expansion}
     A_{\vec{k}+\vec{q},2}^+
   \end{pmatrix}^T
   \vec{\sigma}\cdot[\vec{q}\times\vec{k}]
   \begin{pmatrix}
     A_{\vec{k},1}\\
     A_{\vec{k},2}
   \end{pmatrix}.\\
\end{eqnarray}
It is widespread opinion that in Pauli equation, considered as non-relativistic 
approximation to Dirac equation, the probability density remains 
Schrodinger-like equal to $|\Psi(\vec{r})|^2$, where $\Psi(\vec{r})$ is an 
appropriate wave function. It is stated that for free spin one-half particle 
only probability current density undergoes modification and spin-dependent 
divergenceless term
\begin{equation}
 \frac{1}{2m}\left[
\vec{\triangledown}\times(\Psi^\dagger(\vec{r})\vec{\sigma}\Psi(\vec{r}))\right]
\end{equation}
is added to classical Schrodinger expression \cite{Nowakowski1999,Hodge2014}
\begin{equation}
  \frac{1}{2m}\left(\Psi^\dagger(\vec{r})\vec{\triangledown}\Psi(\vec{r})
   - \Psi(\vec{r})\vec{\triangledown}\Psi^\dagger(\vec{r})
\right).
\end{equation}
As seen from \eqref{eq_Pe_expansion} the derived Pauli probability density 
operator also differs essentially from its canonical counterpart. In order to 
understand the meaning of obtained result let us consider the probability 
density $W(\vec{R})$, that is by the definition
\begin{eqnarray}
 \nonumber
 W_e(\vec{R}) &=&
 \int \langle \varphi(\vec{r})|\hat{P}_e(\vec{R})| \varphi(\vec{r}) \rangle 
   d\vec{r} \\ \nonumber
 &=& \frac{1}{(2\pi)^6} \int
 \langle \varphi(\vec{k} + \vec{q})|\hat{P}_e(\vec{q})| \varphi(\vec{q}) 
 \rangle e^{-i\vec{q}\vec{R}} d\vec{k} d\vec{q}. \\
\end{eqnarray}
Here the averaging is done over single particle functions of the type
\begin{equation}
 |\varphi(\vec{k})\rangle = 
  [\varphi_1(\vec{k})A_{\vec{k},1}^+ + \varphi_2(\vec{k})A_{\vec{k},2}^+]
   |0\rangle
\end{equation}
and it is assumed that
\begin{equation}
 \int \left( |\varphi_1(\vec{k})|^2 + |\varphi_2(\vec{k})|^2\right) d\vec{k}
  = 1.
\end{equation}
Let us consider expansion of $W_e(\vec{R})$ up to the second order in 
$\lambda_C k$. The ``classical'' result is reproduced in the zeroth order in 
$\lambda_C k$.
\begin{eqnarray}
\nonumber
 \delta W_{e,0}(\vec{R}) &\approx& \int \left[
   \varphi_1^*(\vec{k}+\vec{q})\varphi_1(\vec{k}) 
    + \varphi_2^*(\vec{k}+\vec{q})\varphi_2(\vec{k}) \right] \\ \nonumber
  &\times& e^{-i\vec{q}\vec{R}} d\vec{k} d\vec{q} \\
  &=& |\varphi_1(\vec{R})|^2 + |\varphi_2(\vec{R})|^2
\end{eqnarray}
The account of terms of the second order in $\lambda_C k$ leads to the 
following additional spin independent term $\delta W_{e,1}(\vec{R})$ and spin 
dependent term $\delta W_{e,2}(\vec{R})$
\begin{gather}
 \delta W_{e,1}(\vec{R}) = 
   -\frac{1}{8}\lambda_C^2\Delta_R\Phi^\dagger(\vec{R})\Phi(\vec{R}), \\
  \delta W_{e,2}(\vec{R}) = -i\frac{1}{4}\lambda_C^2 
    [\vec{\nabla}_{\vec{r_1}} \times \vec{\nabla}_{\vec{r_2}}]
    \Phi^\dagger(\vec{r_1})\vec{\sigma}\Phi(\vec{r_2})
    \lvert_{\vec{r_1}=\vec{r_2}=\vec{R}}
\end{gather}
Here $\Phi(\vec{R})$ is two component Pauli spinor
\begin{equation}
 \Phi(\vec{R}) = \begin{pmatrix}
     \varphi_1(\vec{R})\\
     \varphi_2(\vec{R})
   \end{pmatrix}.
\end{equation}
The similar terms for the charge and current densities considered as the 
sources in the Maxwell equations were obtained in \cite{PhysRevA.88.032117}. The 
derivation within Lagrangian approach, which is valid up to the second order in 
$\lambda_C k$, was based on a two component Pauli-like equation obtained by 
Foldy-Wouthuysen transformation.

We can use our density operator to incorporate external field into Pauli 
equation. The expectation value of interaction with the classical potential 
field 
$V(\vec{R})$ in electron sector can be written as
\begin{equation}
 \int V(\vec{R})W_e(\vec{R}) d\vec{R}.
\end{equation}
The spin independent terms $\delta W_{e,1}(\vec{R})$ and $\delta 
W_{e,2}(\vec{R})$ generate classical Schrodinger potential term
\begin{equation}
 \int V(\vec{R})\Phi^\dagger(\vec{R})\Phi(\vec{R}) d\vec{R}
\end{equation}
and Darwin term
\begin{equation}
 \frac{1}{8}\lambda_C^2\int 
\Phi^\dagger(\vec{R})\varDelta V(\vec{R})\Phi(\vec{R}) d\vec{R}.
\end{equation}
The spin dependent term generates spin-orbit interaction term
\begin{equation}
 \frac{1}{4}\lambda_C^2\int 
\Phi^\dagger(\vec{R}) \vec{\sigma}
 [\vec{\nabla}V(\vec{R}) \times \hat{\vec{p}}]\Phi(\vec{R}) d\vec{R}.
\end{equation}
Thus within proposed approach, it follows that the external potential 
modification is contained implicitly in the ``proper'' defined density 
probability expression, owning its complicated form to the account for the 
existence of vacuum (filled valence band). It must be stressed that we can 
arrive at the same result within projection operator approach (Foldy-Wouthuysen 
method) in the considered $\lambda_C$ approximation. Within our approach the 
general expression for spin-dependent contribution to probability density 
operator, responsible for Rashba-like interaction, predicts non linear 
dependence on wave vector \cite{PhysRevB.73.113303,PhysRevB.74.193314}. More 
generally, for arbitrary potential its interaction with spin is non-local. 
Really, using \eqref{eq_Pe} we obtain the following term in Pauli Hamiltonian
\begin{eqnarray}
\nonumber
 \Delta H_{SO} &=&  i\lambda_C^2 \int
 \tilde{\Phi}^\dagger(\vec{k}+\vec{q})V(\vec{q})
 \vec{\sigma}\cdot[\vec{q} \times \vec{k}] \tilde{\Phi}(\vec{k})
 d\vec{k}d\vec{q}, \\
\end{eqnarray}
where $\tilde{\Phi}(\vec{k}) = M_{\vec{k}}\Phi(\vec{k})$. In configuration 
space $\Delta H_{SO}$ can be written as
\begin{equation}
 \Delta H_{SO} = \lambda_C^2 \int 
  \Phi^\dagger(\vec{R}_1) \hat{H}_{SO}(\vec{R}_1, \vec{R}_2) \Phi(\vec{R}_2)
   d\vec{R}_1 d\vec{R}_2,
\end{equation}
where non-local operator $\hat{H}_{SO}(\vec{R}_1, \vec{R}_2)$ is
\begin{equation}
 \hat{H}_{SO}(\vec{R}_1, \vec{R}_2) = \int
  K(\vec{R}_1 - \vec{R})\hat{H}_{SO}(\vec{R})K(\vec{R} - \vec{R}_2)
   d\vec{R},
\end{equation}
where $\hat{H}_{SO}(\vec{R})$ is familiar SO operator
$$
\hat{H}_{SO}(\vec{R}) =  \vec{\sigma}[\vec{\nabla}V(\vec{R}) \times 
\hat{\vec{p}}]
$$
and
$$
K(\vec{R}) = \int M_{\vec{k}} e^{-i\vec{k}\vec{R}}.
$$
Alternatively, the arrived result can be viewed upon as the ``smearing'' of the 
wave function. Within both interpretations the physical origin of the effect can 
be attributed to peculiarity of position operator in relativistic Dirac theory. 
As it has been shown in \cite{RevModPhys.21.400} in Dirac theory the position 
operator must be considered as a non-local integral operator in configuration 
space. We can arrived to the same conclusion within Foldy-Wouthuysen approach 
\cite{PhysRevA.84.062124}. It was shown that within our alternative approach it 
can be inferred that position operator components do not commute in general 
\cite{2016arXiv161100900R}. Thus any function of coordinates becomes operator 
and consequently will be ``smeared'' as regards its classical counterpart due 
to emerging uncertainty relation.  The spatial behavior of the ``smearing'' 
kernel $K(\vec{R})$ can be estimated following the line of arguments of 
\cite{Zawadzki2011}. The transformation kernel in 3D case can be written as
\begin{eqnarray}
\nonumber
 K(\vec{R}) &=& \frac{1}{(2\pi)^3} 
  \int \frac{e^{i\vec{k}\vec{R}}}{\sqrt{2\varepsilon_k(\varepsilon_k + 1)}}
   d\vec{k} \\
  &=&  \frac{1}{(2\pi)^3} \int \frac{e^{i\vec{k}\vec{R}}}{\varepsilon_k}
   G(\vec{k}) d\vec{k},
\end{eqnarray}
where $G(\vec{k}) = \sqrt{\varepsilon_k}/\sqrt{2(\varepsilon_k + 1)}$.  This 
function varies very slowly from $G(0) = 0.5$ to $G(\infty) = 1/\sqrt(2) 
\approx 0.707$. Thus, following \cite{Zawadzki2011} we approximate $G(\vec{k})$ 
by some effective constant $G_0$. Using the following expressions 
\cite{Integrals_tbl}
\begin{gather}
 \int_{-\infty}^\infty \frac{e^{itz}}{\sqrt{t^2 + a^2}}dz = 2K_0(az),\\
 \int_0^\infty tJ_0(\beta t)K_0(\alpha\sqrt{t^2+1})dt
  = \frac{K_1(\sqrt{\alpha^2 + \beta^2})}{\sqrt{\alpha^2 + \beta^2}},
\end{gather}
where $K_1(z)$ is the modified Bessel (Mac Donald) function, we obtain
\begin{equation}
 K(\vec{R}) \approx \frac{G_0}{2(2\pi)^2\lambda_C^3}
  \frac{\lambda_C}{R} K_1\left( \frac{R}{\lambda_C} \right).
\end{equation}
Using asymptotic behavior of $K_1(z)$ \cite{Math_func}
\begin{equation}
 K_1(z) \infty \sqrt{\frac{\pi}{2}}\frac{e^{-z}}{\sqrt{z}}
  \left( 1 + O\left( \frac{1}{z} \right) \right).
\end{equation}
We see that the smearing wave function (potential) function have a finite width 
of the order of $\lambda_C$.

In 2D case the kernel describing non-local interaction with external potential 
within the same approximation \cite{Integrals_tbl} is
\begin{eqnarray}
\nonumber
 K(\vec{R}) &=& \frac{1}{(2\pi)^2} \int 
  \frac{e^{i\vec{k}\vec{R}}}{\varepsilon_k}G(\vec{k})d\vec{k} \\ \nonumber
  &\approx& \frac{G_0}{4\pi} \int 
    \frac{J_0(\vec{k}\vec{R})}{\varepsilon_k}d\vec{k}
    =  \frac{G_0}{4\pi} \frac{1}{\lambda_C R} e^{-R/\lambda_C}.
\end{eqnarray}

It must be pointed out that as follows from 
above, the ``relativism'' (Lorentz invariance) per se does not enter presented 
consideration. All consideration is carried out in the fixed frame of reference. 
The critical point is the exsistance of two different types of states with 
positive and negative energies. So, the proposed approach must be valid for the 
analysis of $\vec{k} \cdot \vec{p}$ problems, which are by the definition 
considered in the fixed rest frame of crystal.

Two main aspects of proposed form of probability density operator must be 
underlined As it was shown, all information about ``relativism'' (SO-like and 
Darwin-like terms) is contained in defined probability density operator, while 
external potential is considered as classical. Secondly, the obtained 
functional form of probability operator implies that wave function, obtained as 
the solution of derived two-component Pauli equation, plays subsidiary role. 
The probabilistic interpretation of outcome of experiment is given by averaged 
value of found probability density operator over the corresponding single 
particle wave functions, which are the solutions of Pauli-like equation. 
One more reson for implementation of second quantization scheme will be 
jusified below while considering the time dependence of density operator, which 
is crucial for formulation of continuity equation.

\section{Current density operator}
\label{sec:3}
Following the outlined procedure the charge current density operator in Dirac 
theory is by the definition in “bare” representation is
\begin{eqnarray}
\nonumber
 &\hat{\vec{J}}(\vec{q})& = c \int 
  \Psi^+(\vec{r})\vec{\alpha}\delta(\vec{R}-\hat{\vec{r}})\Psi(\vec{r})
   e^{-i\vec{q}\vec{R}} d\vec{r}d\vec{R} \\ \nonumber
&=& c \int \left[ \begin{pmatrix}
     a_{\vec{k}+\vec{q},1}^+ \\
     a_{\vec{k}+\vec{q},2}^+
   \end{pmatrix}^T \vec{\sigma}
   \begin{pmatrix}
     b_{\vec{k},1}^+ \\
     b_{\vec{k},2}^+
   \end{pmatrix} +
   \begin{pmatrix}
     b_{\vec{k}+\vec{q},1} \\
     b_{\vec{k}+\vec{q},2}
   \end{pmatrix}^T \vec{\sigma}
   \begin{pmatrix}
     a_{\vec{k},1} \\
     a_{\vec{k},2}
   \end{pmatrix} \right] d\vec{k}. \\
\end{eqnarray}
The truncation to electron channel leads to the following expression
\begin{eqnarray}
 \nonumber
 &\hat{\vec{J}}_e(\vec{q})& = c\int
 N_{\vec{k}+\vec{q}} M_{\vec{k}} \lambda_C
   \begin{pmatrix}
     A_{\vec{k}+\vec{q},1}^+ \\
     A_{\vec{k}+\vec{q},2}^+
   \end{pmatrix}^T \vec{\sigma}(\vec{\sigma} \cdot \vec{k})
   \begin{pmatrix}
     A_{\vec{k},1} \\
     A_{\vec{k},2}
   \end{pmatrix} d\vec{k} \\ \nonumber
 &+& c\int
 N_{\vec{k}} M_{\vec{k}+\vec{q}} \lambda_C
   \begin{pmatrix}
     A_{\vec{k}+\vec{q},1}^+ \\
     A_{\vec{k}+\vec{q},2}^+
   \end{pmatrix}^T \left( \vec{\sigma} \cdot (\vec{k}+\vec{q}) 
          \right)\vec{\sigma}
   \begin{pmatrix}
     A_{\vec{k},1} \\
     A_{\vec{k},2}
   \end{pmatrix} d\vec{k}\\
 &=& \hat{\vec{J}}_{1,e}(\vec{q}) + \hat{\vec{J}}_{2,e}(\vec{q})
  + \hat{\vec{J}}_{e,Z}(\vec{q}),
\end{eqnarray}
where
\begin{eqnarray}
 \nonumber
 \hat{\vec{J}}_{1,e}(\vec{q}) &=& c \lambda_C \sum_{s=1,2} \int 
  d\vec{k} \big[ N_{\vec{k}+\vec{q}} M_{\vec{k}}\vec{k} \\
    &+& N_{\vec{k}}M_{\vec{k}+\vec{q}}(\vec{k}+\vec{q})
    \big] A_{\vec{k}+\vec{q},s}^+A_{\vec{k},s},
\end{eqnarray}
\begin{eqnarray}
 \nonumber
 \hat{\vec{J}}_{2,e}(\vec{q}) &=& ic \lambda_C \int d\vec{k}\big[
   N_{\vec{k}+\vec{q}} M_{\vec{k}} \\
  &-& N_{\vec{k}}M_{\vec{k}+\vec{q}} \big]
   \begin{pmatrix}
     A_{\vec{k}+\vec{q},1}^+ \\
     A_{\vec{k}+\vec{q},2}^+
   \end{pmatrix}^T [\vec{k} \times \vec{\sigma}]
      \begin{pmatrix}
     A_{\vec{k},1} \\
     A_{\vec{k},2}
   \end{pmatrix},
\end{eqnarray}

\begin{equation}
\label{eq_J_eZ}
 \hat{\vec{J}}_{e,Z}(\vec{q}) = -ic \lambda_C \int d\vec{k}
  N_{\vec{k}}M_{\vec{k}+\vec{q}}
     \begin{pmatrix}
     A_{\vec{k}+\vec{q},1}^+ \\
     A_{\vec{k}+\vec{q},2}^+
   \end{pmatrix}^T [\vec{q} \times \vec{\sigma}]
      \begin{pmatrix}
     A_{\vec{k},1} \\
     A_{\vec{k},2}
   \end{pmatrix}.
\end{equation}
The first term $\hat{\vec{J}}_{1,e}(\vec{q})$ in the zero order approximation 
in $\lambda_C k$ is
\begin{equation}
 \hat{\vec{J}}_{1,e}(\vec{q}) \approx \frac{1}{2m} \lambda_C \sum_{s=1,2} \int 
  d\vec{k} [2\vec{k}+\vec{q}] A_{\vec{k}+\vec{q},s}^+A_{\vec{k},s}.
\end{equation}
It is easy to recognize ``classical'' result in this term after averaging over 
corresponding wave functions
\begin{eqnarray}
\nonumber
 &&\int \langle \varphi(\vec{r})|\hat{\vec{J}}_{1,e}(\vec{q})| \varphi(\vec{r}) 
\rangle d\vec{r} \\
&=& -\frac{i}{2m}\left( \Psi^\dagger(\vec{R})\vec{\triangledown}\Psi(\vec{R})
   - \Psi(\vec{R})\vec{\triangledown}\Psi^\dagger(\vec{R}) \right) .
\end{eqnarray}

The last divergenceless ($\vec{q} \cdot \hat{\vec{J}}_{e,Z}(\vec{q}) \equiv 0$) 
term $\hat{\vec{J}}_{e,Z}(\vec{q})$ represents so called spin magnetization 
current. The necessity of its presence was established long ago by 
G.Breit\cite{PhysRev.53.153}. The addition of this current in considered 
approximation to the classical Schrodinger expression has been justified in 
\cite{Nowakowski1999,PhysRevA.88.032117}. The necessity for accounting of this 
term in current is supported by the fact that the Zeeman term arises due to it 
in Pauli Hamiltonian. Really, by choosing
$$
\vec{A}(\vec{R}) = \frac{1}{2}[\vec{B} \times \vec{R}]
$$
for constant magnetic field $\vec{B}$, the corresponding term in expression for 
expectation value of interaction with external electromagnetic field is
\begin{eqnarray}
\nonumber
 H_{Z,1} &=& \frac{e}{c}\int \vec{A}(\vec{R}) \langle 
\hat{\vec{J}}_{e,Z}(\vec{q}) \rangle e^{-i\vec{q}\vec{R}} d\vec{q} d\vec{R} \\
&=& i\frac{e}{2c}\int [\vec{B} \times \vec{\nabla}\delta(\vec{q})]
 \langle \hat{\vec{J}}_{e,Z}(\vec{q}) \rangle d\vec{q}.
\end{eqnarray}
After some manipulation using well-known equality
\begin{equation}
 [\vec{A}\times\vec{B}][\vec{C}\times\vec{D}]
 = (\vec{A}\vec{C})(\vec{B}\vec{D})-(\vec{A}\vec{D})(\vec{B}\vec{C})
\end{equation}
and expression \eqref{eq_J_eZ}, the expectation value of interaction with 
external electromagnetic field can be presented as the Zeeman term
\begin{equation}
  H_{Z,1} = \frac{e}{2c}\int 
  \bigg\langle \frac{\vec{B}\vec{\sigma}}{m \varepsilon_k } \bigg\rangle
   d\vec{k}.
\end{equation}
Note that in relativistic mechanics $\varepsilon_k=\sqrt{1 - v^2/c^2}$. This 
result means that in Zeeman term the rest mass in the expression for spin 
magnetic moment of Pauli electron must be replaced by expectation value of 
energy-dependent Lorentz mass in the corresponding quantum state. Compare this 
result with \cite{PhysRevD.3.1728}. The $\lambda_C^2$ expansion of this 
expression up to the second order produces the relativistic correction term to 
the Zeeman interaction $g\vec{\mu}\vec{B}$
\begin{equation}
 H_{Z,1} \approx -\frac{e}{4mc}\lambda_C^2 \int
   \langle k^2 \vec{B}\vec{\sigma}\rangle d\vec{k}.
\end{equation}
This addition to intrinsic magnetic moment due to the $k^4/8m^3 c^2$ term in 
expansion of relativistic kinetic energy was considered in \cite{Meister1980}. 
This dependence of mass on velocity must be accounted for if spin-orbit 
interaction is taken into account, as its contribution to single particle 
Hamiltonian is of the same order in $\lambda_C^2$.

Nevetheless this result cannot be considered as final. The point is that there 
is additional spin-dependent term $\hat{\vec{J}}_{2,e}(\vec{q})$, depending also 
on $\vec{B}(\vec{r})$, but not explicitly on $\vec{A}(\vec{r})$. Thus it must 
be added to written above Zeeman term.
\begin{eqnarray}
 \nonumber
 H_{Z,2} &=& \frac{e}{2mc} \int 
 \langle [\vec{k} \times \vec{\sigma}]
  (N_{\vec{k}+\vec{q}}M_{\vec{k}} - N_{\vec{k}}M_{\vec{k}+\vec{q}}) \rangle \\
   \nonumber
   &\times& [\vec{B} \times \delta'(\vec{q})] d\vec{k}d\vec{q} \\
   \nonumber
   &=& -\frac{3}{4}\lambda_C^2 \Big\langle
    \frac{(\vec{B}|\vec{k})(\vec{k}|\vec{\sigma}) - (\vec{B}|\vec{\sigma})k^2}
    {\varepsilon_k^2(1 + \varepsilon_k)} \Big\rangle \\
   &=& -\frac{e}{2c} \Big\langle 
       \frac{\Delta\hat{\vec{S}}_{FW}\vec{B}}{m\varepsilon_k} \Big\rangle .
\end{eqnarray}
It is easy to recognize in the term $\Delta\hat{\vec{S}}_{FW}$ the one
appearing in spin operator of Foldy-Wouthuysen theory. In accord with our 
no-pair approximation we retain only even part $\hat{\vec{S}}_{FW}^{even}$ of 
defined by them spin operator $\hat{\vec{S}}_{FW}$ 
\cite{PhysRevA.89.052101,PhysRevA.88.022119} which is in our notation
\begin{equation}
\hat{\vec{S}}_{FW}^{even} = \frac{1}{2}\hat{\vec{\sigma}}
 - \frac{1}{2}\lambda_C^2 \frac{\vec{k} \times 
[\hat{\vec{\sigma}}\times\vec{k}]}{\varepsilon_k(1 + \varepsilon_k)} .
\end{equation}
For massless situation describing charged ``neutrinos'' (graphene case) the 
Zeeman term is
\begin{gather}
 H_{Z,1} = \frac{1}{2}e\int \Big\langle
  \frac{[\vec{B}\times\vec{\sigma}]}{|\vec{k}|^2} \Big\rangle d\vec{k}, \\
 H_{Z,2} = \frac{1}{2}e\int \Big\langle
 \frac{[\vec{B}\times\vec{k}][\vec{k}\times\vec{\sigma}]}{|\vec{k}|^3}
  \Big\rangle d\vec{k} .
\end{gather}

\section{Continuity equation}
\label{sec:4}
In the first quantization scheme the probability density current in Dirac 
theory \mbox{$\hat{\vec{J}}(\vec{k},t) = 
c\vec{\alpha}\exp(-i\vec{k}\hat{\vec{r}}(t))$} is connected with the 
probability density \mbox{$\hat{P}(\vec{k},t) = 
\exp(-i\vec{k}\hat{\vec{r}}(t))$} via continuity equation
\begin{equation}
\label{eq_Continuity}
\frac{\partial \hat{P}(\vec{k},t)}{\partial t} = 
 i [\hat{H}, \hat{P}(\vec{k},t)] = -i\vec{k}\hat{P}(\vec{k},t).
\end{equation}
The time dependence of $\hat{P}(\vec{k},t)$ in the Heisenberg representation 
is determined in SQM by the time dependence of creation/annihilation operators.
In the free of external perturbations Dirac problem
\begin{gather}
 A_{\vec{k}}(t) = e^{-imc^2 \varepsilon_k t}A_{\vec{k}}, \\
 A^+_{\vec{k}}(t) = e^{imc^2 \varepsilon_k t}A^+_{\vec{k}}.
\end{gather}
Thus, the time derivative of \eqref{eq_Pe} is
\begin{widetext}
\begin{eqnarray}
\nonumber
 \frac{\partial \hat{P}_e(\vec{q},t)}{\partial t} &=& imc^2 \sum_{s=1,2} \int
  (\varepsilon_{k+q} - \varepsilon_k)[N_{\vec{k}+\vec{q}}N_{\vec{k}}
  + M_{\vec{k}+\vec{q}}M_{\vec{k}}\lambda_C^2
    \left( (\vec{k}+\vec{q}) \cdot \vec{k} \right) ]
    A^+_{\vec{k}+\vec{q},s}(t) A_{\vec{k},s}(t) d\vec{k} \\
 &-& \frac{1}{m} \sum_{s=1,2} \int
 M_{\vec{k}+\vec{q}}M_{\vec{k}} (\varepsilon_{k+q} - \varepsilon_k)
  \begin{pmatrix}
     A_{\vec{k}+\vec{q},1}^+(t) \\ \nonumber
     A_{\vec{k}+\vec{q},2}^+(t)
   \end{pmatrix}^T
   \vec{\sigma}\cdot[\vec{q}\times\vec{k}]
      \begin{pmatrix}
     A_{\vec{k},1}(t)\\
     A_{\vec{k},2}(t)
   \end{pmatrix}d\vec{k} .
\end{eqnarray}
\end{widetext}
The ``Zeeman'' current
$\hat{\vec{J}}_{e,Z}(\vec{q},t)$ being divergenceless does not contribute to the 
continuity equation. Nevertheless, it must be retained because it is responsible 
for Zeeman interaction.

As the Dirac probability current density $\vec{\alpha}\delta(\vec{R}- 
\hat{\vec{r}})$ is not affected by switching on of electromagnetic field, the 
famous term $\vec{A}(\vec{r})\Psi^*(\vec{r})\Psi(\vec{r})$ can not appear in our 
non-relativistic expression for Pauli current. Nevertherless, this term is 
essential for the concerving of the gauge invariance. So, the question arises 
where do we lost it? The point is that we have not taken into account changing 
of equations of motion for annihilation/creation operators due to the Peierls 
substitution \mbox{$\vec{p} - \tfrac{e}{c} \vec{A}(\vec{r})$}, when 
electromagnetic field is switched on. It can be shown in our approach, that the 
following term appears accordingly in single particle Pauli equation (in the 
first order in parameter $\lambda_C k$)
\begin{eqnarray}
\nonumber
 \Delta H_A &=& e\lambda_C \sum_s \int 
  A^+_{\vec{k},s} \vec{q}\vec{A}(\vec{k} - \vec{q}) A_{\vec{q},s} 
    d\vec{k}d\vec{q} \\ \nonumber
 &-& \frac{1}{2}e\lambda_C \sum_{s,s'} \int 
 A^+_{\vec{k},s}\{\vec{\sigma}\vec{B}(\vec{k} - \vec{q})\}_{s,s'}A_{\vec{q},s'}
    d\vec{k}d\vec{q} .\\
\end{eqnarray}
The time dependence of annihilation operators with account to this addition is
\begin{eqnarray}
\nonumber
 \frac{\partial A_{\vec{q},s}(t)}{\partial t} &=& -i mc^2 \varepsilon_k 
     A_{\vec{q},s}(t) \\ \nonumber
 &-& ie\lambda_C \sum_s \int A_{\vec{q},s}(t) \vec{q}\vec{A}(\vec{k} - \vec{q})
  d\vec{q} \\ \nonumber
 &+& \frac{1}{2} e\lambda_C \sum_{s'} \int
  \{\vec{\sigma}\vec{B}(\vec{k} - \vec{q})\}_{s,s'}A_{\vec{q},s'}(t) d\vec{q}.\\
\end{eqnarray}
The last term describing interaction of spin with magnetic field (Zeeman term) 
was discussed in detail for constant magnetic field above. Thus we will be 
interested in the second term. It is streightfarward to 
show that this correction can be rewritten in the form
\begin{eqnarray}
\nonumber
 &\Delta&\frac{\partial \hat{P}(\vec{q},t)}{\partial t} = 
  i\vec{q}\hat{\vec{I}}_{e,\vec{A}}(\vec{q},t) \\ \nonumber
 &=& iec \lambda_C \sum_s \int
   A^+_{\tilde{\vec{k}},s}(t) \vec{q}\vec{A}(\tilde{\vec{k}}-\vec{k}-\vec{q})
   A_{\vec{k},s}(t)d\tilde{\vec{k}}d\vec{k}. \\
\end{eqnarray}
It is easy to verify that in position space this term averaged over single 
electron wave function is just the sought 
\mbox{$\tfrac{e}{m}div\vec{A}(\vec{r}) \Psi^*(\vec{r})\Psi(\vec{r})$}. It must 
be stressed that the appearance of such addition to $\hat{P}_e(\vec{q},t)$ time 
derivative can not be compensated in continuity equation by the proposed 
truncated $\hat{\vec{J}}(\vec{q},t)$. In order to preserve the constancy of the 
total probability \mbox{$P=\int \langle \hat{P}(\vec{R}) \rangle d\vec{R} 
\equiv 1$} we must complement continuity equation by the source term
\begin{equation}
 \frac{\partial \hat{P}_e(\vec{q},t)}{\partial t} - 
  i\vec{q}\hat{\vec{J}}_e(\vec{q},t)
 =  i\vec{q}\hat{\vec{I}}_{e,\vec{A}}(\vec{q},t).
\end{equation}
The appearance of the source term induced by 
$\vec{q}\hat{\vec{I}}_{e,\vec{A}}(\vec{q},t)$, signifies that probability 
current operator derived directly from Dirac current is not conserved when 
external electromagnetic field is applied to the system. Rearranging this term 
to the left side we have
\begin{equation}
 \frac{\partial \hat{P}_e(\vec{q},t)}{\partial t} 
 - i\vec{q}\left( \hat{\vec{J}}_e(\vec{q},t) + 
\hat{\vec{I}}_{e,\vec{A}}(\vec{q},t) \right) = 0.
\end{equation}
Which is now in the form of the standard sourceless continuity equation if we 
redefine current as
\begin{equation}
 \hat{\vec{J}}_{e,total}(\vec{q},t) =
 \hat{\vec{J}}_e(\vec{q},t) + \hat{\vec{I}}_{e,\vec{A}}(\vec{q},t) .
\end{equation}
The source term $\hat{\vec{I}}_{e,\vec{A}}(\vec{q},t)$ is a material property 
and must vanish outside the sample. It is analogous to the contribution to the 
charge density from divergence of the polarization in electrodynamics of 
continuous matter \cite{Electrdyn_cont_mat}. Here the SQM vacuum plays the role 
of ``polarized'' perturbated medium. Such essential physical difference in the 
origin of these two contributions provides the possibility to 
impose different constrains on their spatial and temporal behavior. For 
example, it is of importance in the theory of superconductivity, where these 
two contributions to the current are treated on different grounds. The first 
paramagnetic contribution to the current is set to zero assuming rigidity of 
macroscopic superconductor wave function or in another words proposing the 
stability of Bose vacuum of Couper pairs under external perturbation. The second 
diamagnetic term, dependent on vector potential, is considered to be different 
from zero and describes the penetration of external magnetic field into 
superconducting media. It must be noted once more, that obtained within proposed 
approach simple form \mbox{$\tfrac{e}{mc}div\vec{A}(\vec{r}) 
\Psi^*(\vec{r})\Psi(\vec{r})$} of material current is valid only in the zeroth 
approximation in $\lambda_C k$. In general, the expression for this current 
depends non-locally on the vector potential in position space. There is a 
different approach to the derivation of probability current density outlined in 
classic textbook of Landau and Lifshitz \cite{Non_rel_quant_mech}. It is based 
on vector potential variation of phenomenologically written down approximate 
Pauli equation for electron. While their approach gives the same expression for 
$\hat{\vec{J}}_{e,total}$, it does not reveal the essential physical difference 
in the origin of these two contributions.

If the external scalar potential perturbation $V(\vec{r})$ is applied to the 
system, the equations of motion for creation/annihilation operators are 
governed in electron channel by the Hamiltonian
\begin{equation}
 \hat{H}_e = \hat{H}_{0,e} + \hat{H}_{int,eV},
\end{equation}
where $\hat{H}_0$ - the Hamiltonian of free Dirac problem and 
$\hat{H}_{int,eV}$ is of the form
\begin{widetext}
\begin{eqnarray}
\nonumber
\hat{H}_{int,eV} &=& \sum_{s=1,2}\int V(\vec{q})\left[ 
   N_{\vec{k}+\vec{q}}N_{\vec{k}} + M_{\vec{k}+\vec{q}}M_{\vec{k}}
    ((\vec{k}+\vec{q}) \cdot \vec{k}) \right]
    A_{\vec{k}+\vec{q},s}^+A_{\vec{k},s}d\vec{k}d\vec{q} \\
&+& i\lambda_C^2 \int V(\vec{q})M_{\vec{k}+\vec{q}}M_{\vec{k}}
 \begin{pmatrix}
     A_{\vec{k}+\vec{q},1}^+(t) \\ \label{eq_Hint_eV}
     A_{\vec{k}+\vec{q},2}^+(t)
   \end{pmatrix}^T
   \vec{\sigma}\cdot[\vec{q}\times\vec{k}]
      \begin{pmatrix}
     A_{\vec{k},1}(t)\\
     A_{\vec{k},2}(t)
   \end{pmatrix}d\vec{k} .
\end{eqnarray}
\end{widetext}
Following the procedure outlined above, it is easily verified that the terms of 
the zeroth order in $\lambda_C k$ do not contribute to 
$\tfrac{\partial}{\partial t}P(\vec{q},t)$. Only the second order term, which 
is spin dependent, remains
\begin{eqnarray}
\nonumber
&&\delta \frac{\partial}{\partial t}\hat{P}(\vec{q},t) 
 = i\vec{q}\vec{I}_{e,V}(\vec{q},t) \approx \\
\nonumber
&\approx&
\frac{\lambda_C^2}{2}\int\int V(\tilde{\vec{q}}) A_{\vec{Q},s}^+(t)
 \vec{q}[\tilde{\vec{q}} \times \vec{\sigma}]_{s,s'}
  A_{\vec{Q}-\vec{q}-\tilde{\vec{q}},s'}(t)d\vec{Q}d\tilde{\vec{q}}\\
\end{eqnarray}
In configuration representation, averaged over two-component electron Pauli 
spinors $\Phi(\vec{r})$, this contribution leads to the appearance in continuity 
equation additional material current $\hat{\vec{I}}_{e,V}(\vec{r},t)$ of the 
form
\begin{equation}
\hat{\vec{I}}_{e,V}(\vec{r},t) \approx
 \frac{1}{2}\lambda_C^2[\vec{\nabla}V(\vec{r}) \times \Phi(\vec{r},t)^\dagger
  \vec{\sigma} \Phi(\vec{r},t)].
\end{equation}
Compare expression for $\hat{\vec{I}}_{e,V}(\vec{r},t)$ with 
\cite{PhysRevA.88.032117,Hodge2014}. It must be underlined that coincidence with 
their results occurs only for this expansion, obtained from general one 
\eqref{eq_Hint_eV} up to the second order in $\lambda_C k$. The expressions in 
the cited papers are valid only within this approximation, while within proposed 
approach the only constrain is imposed upon is the potential strength. It must 
be weak enough to create real electron/positron (hole) pairs. One more striking 
difference lies in the approach to derivation of this term. ``Classical'' 
consideration outlined in Landau and Lifshitz textbook \cite{Non_rel_quant_mech} 
is based on variation of applied external vector potential. It requires the 
single particle Hamiltonian, written up to the second order in $\lambda_C k$, 
accounting for classical spin-orbit interaction and Peierls substitution 
\mbox{$\vec{p} \rightarrow \vec{p} + e/c \vec{A}(\vec{r})$}. As it follows from 
the considerations presented above, the account for vector 
potential, while deriving this term, is superfluous. This current term is solely 
due to external potential. Moreover, as in the case of redefinition of 
probability current under electromagnetic action, we are to expect that 
different spatial and temporal conditions can be imposed on 
$\hat{\vec{I}}_{e,V}(\vec{q},t)$ and $\hat{\vec{J}}_{e}(\vec{q},t)$. Such 
situation (as in superconductivity) requires e.g. the stability of bulk state 
under application of electric field and existence of dissipationless surface 
currents. Exactly this situation is realized in topological insulators.

\section{Summary}
We proposed SQM approach for construction of probability and probability current 
density operators for single particle Pauli-like equation as an alternative to 
Foldy-Wouthuysen consideration. We partitioned the Hamiltonian (external 
perturbation accounted for) and operators of interest into the part acting 
within electron (hole) stationary states and resudial part responsible for pair 
creation/annihilation processes. The effect of resudial part is fully neglected 
in the present paper, assuming that considered external perturbations are weak 
enough and does not depend on time. The semirelativistic probability and 
probability current operators, defined within such approximation, demonstrate 
Gordon-like structure \cite{Gordon1928}, splitting into spin dependent and 
spin-independent parts. Proposed approach allows to go beyond the commonly used 
$1/c^2$ approximation. Thus, for example, the defined probability operator 
predict non-linear dependence on particle momentum for Rashba-like interaction. 
It was also inferred that in Zeeman term the rest mass in the expression for 
spin magnetic moment of Pauli electron must be replaced in general by 
expectation value of energy-dependent Lorentz mass in the corresponding quantum 
state. The application of external perturbation leads to violation of simple 
form of continuation equation. It is shown that within proposed approach the 
continuity equation is valid in general form with sources, dependent on external 
fields. The source terms can be represented as a divergence of some 
``material'' currents, in full analogy with the description of electro-magnetic 
response in the theory of continuous matter 
\cite{Electrdyn_cont_mat,PhysRevA.88.032117,Hodge2014}. 

Based on semi-relativistic similarity of Dirac problem and multicomponent 
$\vec{k} \cdot \vec{p}$  Hamiltonians, we intend to extend the considered 
approach for the derivation of single quasipaticle probability operators in 
semiconductors.

\end{document}